\documentclass[twoside,fleqn]{ActaStyle}
       \usepackage[dvips]{graphicx}
       \usepackage{times,cite}
   \def\authorheading{V.V. Skokov, V.D. Toneev}
   \def\shorttitle{Dilepton production from ...}
\def\lsim{\mathrel{\rlap{
\lower4pt\hbox{\hskip-3pt$\sim$}}
    \raise1pt\hbox{$<$}}}     %less than approx. symbol
\def\gsim{\mathrel{\rlap{
\lower4pt\hbox{\hskip-3pt$\sim$}}
    \raise1pt\hbox{$>$}}}     %greater than or approx. symbol
   \begin{document}
   \pagerange{1}{7}   %(paper has 7 pages)
   \title{DILEPTON PRODUCTION \\ FROM HYDRODYNAMICALLY EXPANDING FIREBALL }
   \author{V.~V.~Skokov\email{vvskokov@theor.jinr.ru}$^{*\dag}$,
   V.~D.~Toneev$^{*\dag}$}
             {$^*$Bogoliubov Laboratory of Theoretical Physics, \\
             Joint Institute for Nuclear Research, 141980, Dubna,
             Russia\\
             $^\dag$Gesellschaft f\"ur Schwerionenforschung,
          64291 Darmstadt, Germany
              }

   \abstract{ A hybrid model is put forward for describing relativistic heavy
ion collisions. The early interaction stage responsible for
entropy creation is calculated within the transport  Quark-Gluon
String Model resulting in an initial state. The passage to
subsequent isoentropic expansion proceeds with exact account for
all conservation laws. Relativistic 3D hydrodynamics with the
mixed phase equation of state is applied to this expansion stage.
Essential differences in evolution of a fireball, described within
this model and one for the Bjorken regime, are noted. The
in-medium modified $e^+e^-$ spectra are studied  and confronted
with the recent CERES/NA45 data for $8\%$ central Pb+Au collisions
at the bombarding energy 158 AGeV. }

%
%   \section{Introduction}
%
%   text
%
%   %%%%%%%%%%%%%%%%%%%%%%%%%%%%%%%%%%%%%%%%%%%%%%%%%%%%%%%%%%%%%%%%%%

 % \section{Introduction}
Electromagnetic probes and, in particular, di-electrons play an
exceptional role providing almost undisturbed information on a
state of highly compressed and hot nuclear matter (fireball)
formed in relativistic heavy ion collisions. Generally, dilepton
yield depends on both global properties of matter constituents
(hadrons and/or quarks, gluons) defined by the equation of state
and  also on individual constituent properties related to their
in-medium modification. Such a modification has been observed by
the CERES Collaboration in the analysis of the $e^+e^-$ invariant
mass spectra from central Pb+Au collisions~\cite{CERES}. The
measured dilepton
 excess in the range of invariant dilepton masses $0.2
\lsim M \lsim 0.7$ GeV may be interpreted  in terms of a strong
in-medium $\rho$-meson modification (see review
articles~\cite{CB99,RW00}). Many papers are devoted to the
analysis of these CERES data but mainly two scenarios of hadron
modification are available on the market: Consideration based on
the Brown-Rho  scaling hypothesis~\cite{BR91} assuming a dropping
$\rho$ mass and that for a strong broadening as found in the
many-body approach by Rapp and Wambach~\cite{RW00,RW1}.
Unfortunately, a very simplified dynamical treatment is usually
applied which obscures the origin of the model agreement or
disagreement with experiment.

In this paper, we make emphasis on nuclear interaction dynamics to
clarify which states in the main contribute to the observed
dilepton yield as well as which are the model parameters used. We
confront our model calculations  with the recent CERES/NA45
results for $8\%$ central Pb+Au collisions at the bombarding
energy 158 AGeV~\cite{CERES/NA45}.

We treat dynamics of heavy ion collisions in terms of a hybrid
model where the initial interaction stage is described by the
transport Quark Gluon String Model (QGSM)~\cite{QGSM}.

\begin{figure}
\begin{center}
  \includegraphics[width=6cm]{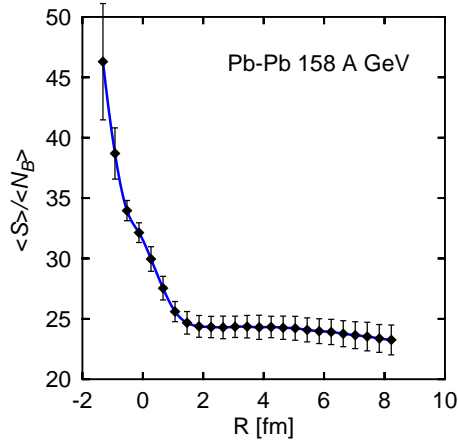}
 \caption{Average entropy $S$ per average baryon charge $Q_B$ for participants versus
 relative distance between the centers of two colliding Pb nuclei at $E_{lab}=$158 $A$GeV.}
 \label{fig:1}
\end{center}
\end{figure}

 In Fig.\ref{fig:1}, the temporal dependence of the ratio of
 entropy $S$ to the total
participant baryon charge $Q_B$ is shown for Pb+Pb collisions at
the impact parameter $b=2.5$ fm and bombarding energy 158 $A$GeV
evaluated within QGSM. Being calculated on a large 3D grid, this
ratio is less sensitive to particle fluctuation  as compared to
entropy itself. Small values of $Q_B$ at the very beginning of
interaction result in large values of the $S/Q_B$ ratio. It is
clearly seen that  this ratio is practically kept constant for
$R\gsim 1.5$ fm, i.e., a little bit later on the moment when two
nuclei completely overlap ($R=0$). The  stage for $R\gsim 1.5$ fm
may be treated as isoentropic expansion.

\begin{figure}[h]
\begin{center}
 \includegraphics[width=6.cm]{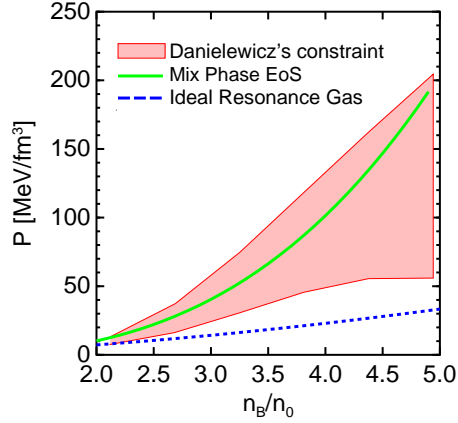}
\caption{Baryon density dependence of pressure at the zero
 temperature. The mixed phase model results are plotted by the solid
 line, the dashed one corresponds to the ideal gas EoS. The shaded
 area is the constraint obtained by Danielewicz {\it et al.}~\cite{Dan}.}
\label{fig:2}
\end{center}
\end{figure}

 So the subsequent stage starts from  $R\approx 1.5$ fm, which corresponds to the
 kinetic evolution time moment  $t_{kin}= 1.8$ fm/$c$ and
 evolves further as a  isoentropic expansion. This stage is described
 within the relativistic 3D hydrodynamics.  A full set of
 hydrodynamic equations is solved  numerically  by the  FCT-SHASTA
 method~\cite{ST05}  with the mixed phase Equation of
State (EoS)~\cite{TFNFR04}. This thermodynamically consistent EoS
uses the modified Zimanyi mean-field interaction for hadrons which
reproduces properties of the saturated nuclear matter at the
normal density $n_B=n_0$ and is in good agreement with the
Danielewicz constraints to EoS~\cite{Dan} (see Fig.\ref{fig:2}).
This mixed phase EoS also includes the interaction between hadron
and quark-gluon phases, which results in a cross-over
deconfinement phase transition.
 In addition to the model prescription in Ref.~\cite{TFNFR04}, the hard
 thermal loop term was
self-consistently added to the interaction of quarks and gluons to
get the correct asymptotics at $T>>T_c$ and to provide reasonable
agreement between the model results and lattice QCD calculations
at finite temperature $T$ and chemical potential
$\mu_B$~\cite{latEoS}.

\begin{figure}[th]
\begin{center}
 \includegraphics[width=6.cm]{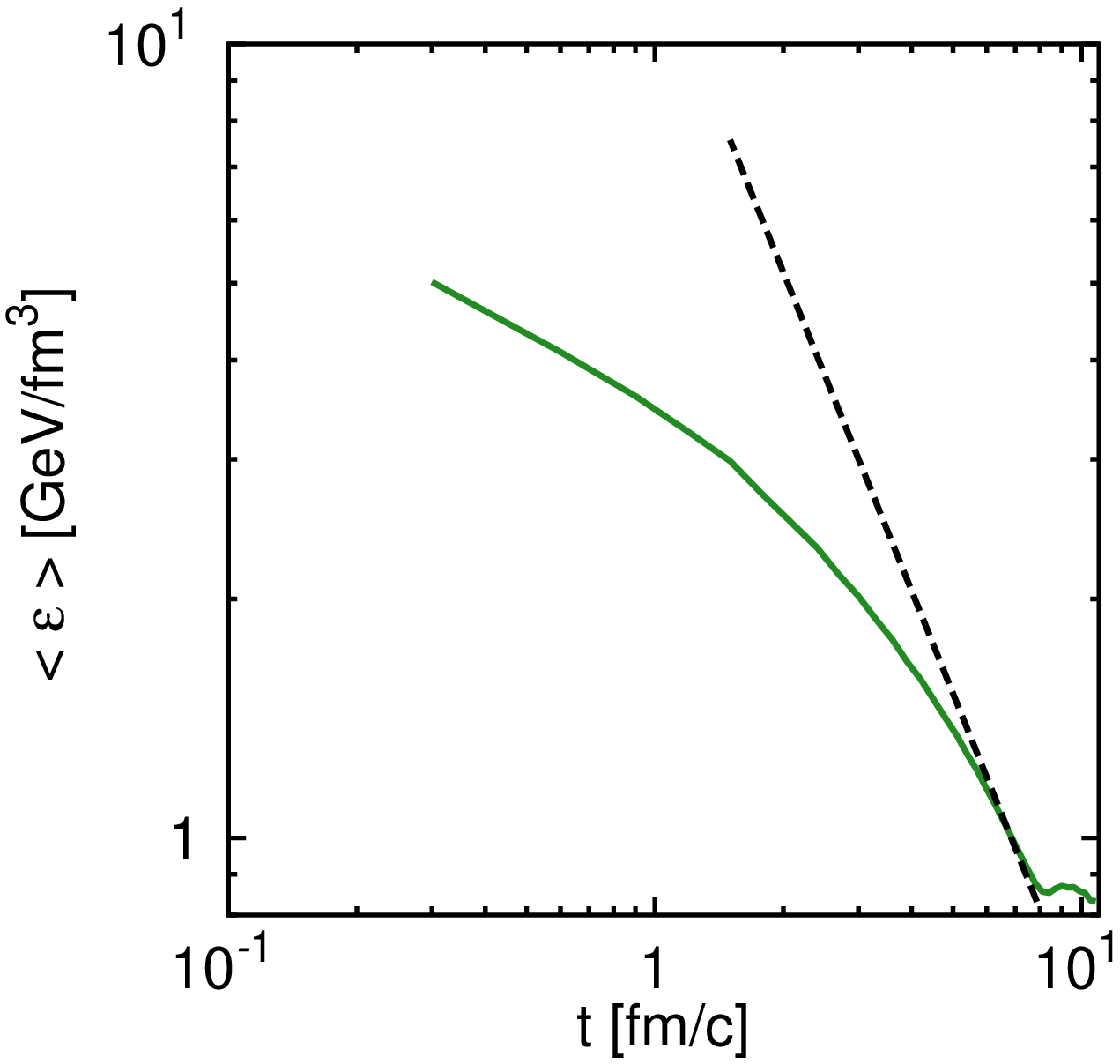}\vspace*{4mm}
  \includegraphics[width=6.cm]{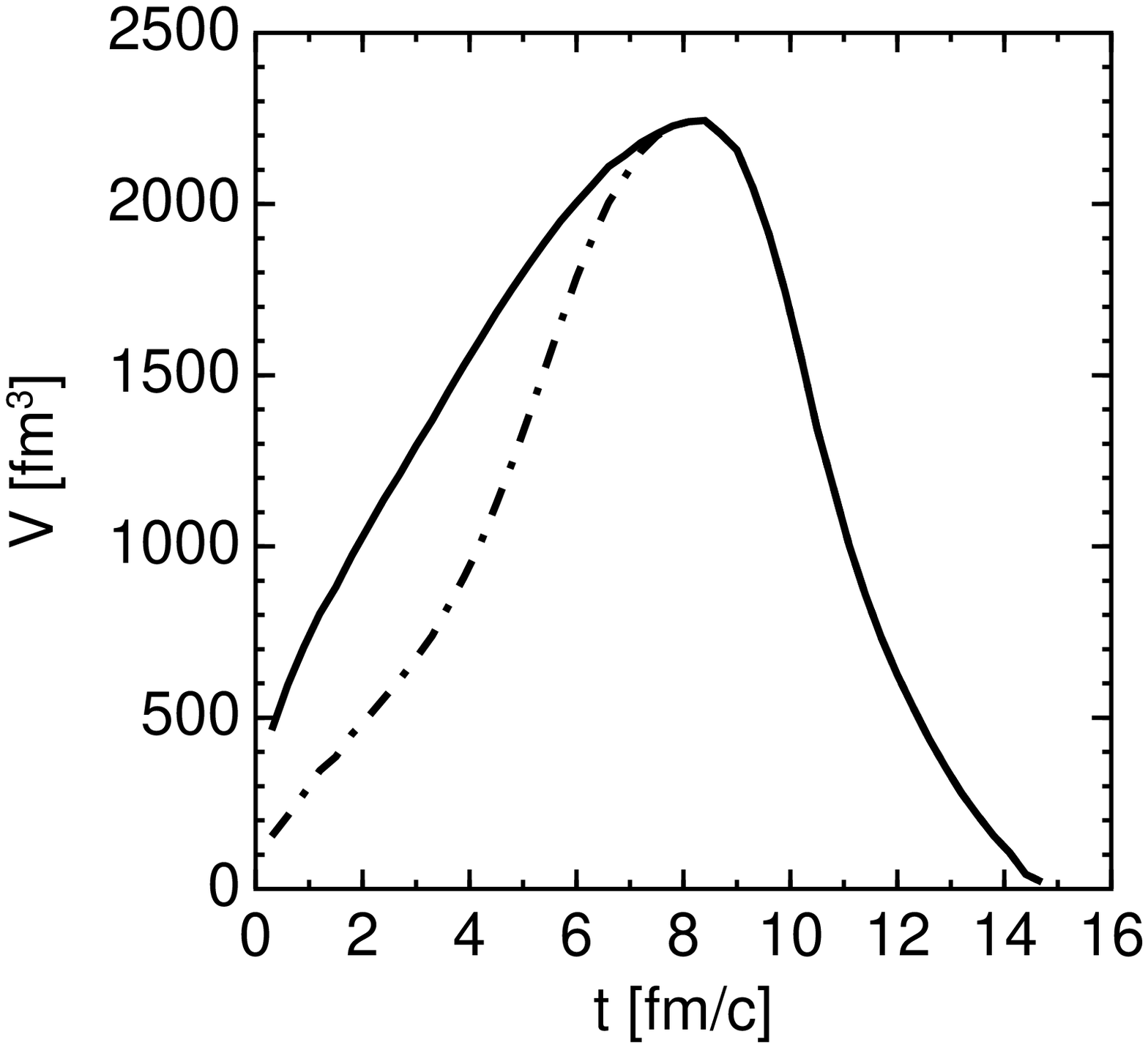}
 \caption{The average energy  density (the left panel) and  volume
 (the right panel) of the expanding  fireball formed
  in central Pb+Pb collisions. The dotted line (in arbitrary
  normalization) shows the Bjorken solution~\cite{B83} for 1D
expansion with the ultrarelativistic ideal gas EoS and
  the dot-dashed line corresponds to the account for hadronic phase cells
  only (see the text).}
  \label{fig:3}
\end{center}
\end{figure}

To proceed from kinetics to hydrodynamics  keeping exactly the
conservation laws, we evaluate locally on the 3D grid the
energy-momentum tensor, baryon density and velocity profile at the
time moment $t_{kin}=1.8$ fm/$c$ and then treat these quantities
as an initial (t=0) state of a fireball for its subsequent 3D
hydrodynamic evolution. The temporal dependence of the average
energy density  is presented  in Figs.\ref{fig:3} for the
expansion stage. The average quantity is defined as
$$ \bar{\varepsilon}(t) \equiv
\bar{\varepsilon}=\int d^3x \ \varepsilon (t,x) \ w(t,x) \ / \
\int d^3x \ w(t,x),$$ where $w(t,x)$ is its time-space
distribution.

As  is seen, the   initial energy density
 is quite high, $\bar{\varepsilon}(t\approx 0)\sim 5.5 \ GeV/fm^3$.
The appropriate compression ratio $\bar{n}_B(0)/n_0$ reaches about
4.2. Note that these values characterize the initial locally
equilibrium state of the fireball and are noticeably lower than
the peak-values estimated in nonequilibrium models with averaging
over some limited spacial volume~\cite{FNT98}. Average energy
density falls down in time and its behavior at the SPS energy
remarkably differs from that for the  Lorentz invariant Bjorken
expansion regime (see the dashed line in Fig.\ref{fig:3}). The
scaled Bjorken solution of 1D hydrodynamics~\cite{B83} is not
applicable to the first moments of hydrodynamic expansion and more
corresponds to the asymptotic expansion regime. However, if the
same initial energy density is assumed for the Bjorken expansion,
one can see that the fireball life time becomes very short. In
addition, this simple solution does not reproduce some flattening
of the time dependence of the energy density  observed for $t\gsim
8$ fm/c. The origin of this effect is demonstrated in the right
side of Fig.\ref{fig:3}.

In our consideration, the hydrodynamic evolution of a cell ends at
the freeze-out defined by the condition that the local energy
density (including 6 neighbor cells) is below a certain value
$\varepsilon_i< \varepsilon_{fr}=0.3$ GeV/fm$^3$. This condition
may be fulfilled locally at the periphery of the expanding system
from the very beginning of expansion and works continuously during
the whole evolution. The frozen cell will not participate in a
further dynamical process but contribute to the "hadron decay
cocktail". The system volume $V$ presented in Fig.\ref{fig:3} is
defined as the sum over all cells with nonvanishing energy density
$\varepsilon_i$ besides frozen cells. As is seen, the system
volume naturally starts to grow with expansion but later on the
competition between expansion and freeze-out effect is getting
stronger and so $V$ drops for $t\gsim 8$ fm/c. Certainly, this
effect is absent in the Bjorken regime. According to the mixed
phase model~\cite{TFNFR04}, in every state generally there is a
homogeneous mixture of free quarks/gluons and bound quarks
(hadrons) with quark densities $\rho_{quarks}^{Q+G}$ and
$\rho_{quark}^{H}$, respectively. One may conditionally define the
hadronic phase as $\rho_{quark}^{H}>\rho_{quarks}^{Q+G}$. As
follows from Fig.\ref{fig:3} (the dot-dashed line), the admixture
of the quark/gluon phase is negligible for the expansion time
$t\gsim 4$ fm/c which roughly corresponds to the average
temperature of the system $\bar{T}(t\approx 4)\approx 180$ MeV.

\begin{figure}[thb]
\begin{center}
  \includegraphics[width=6cm]{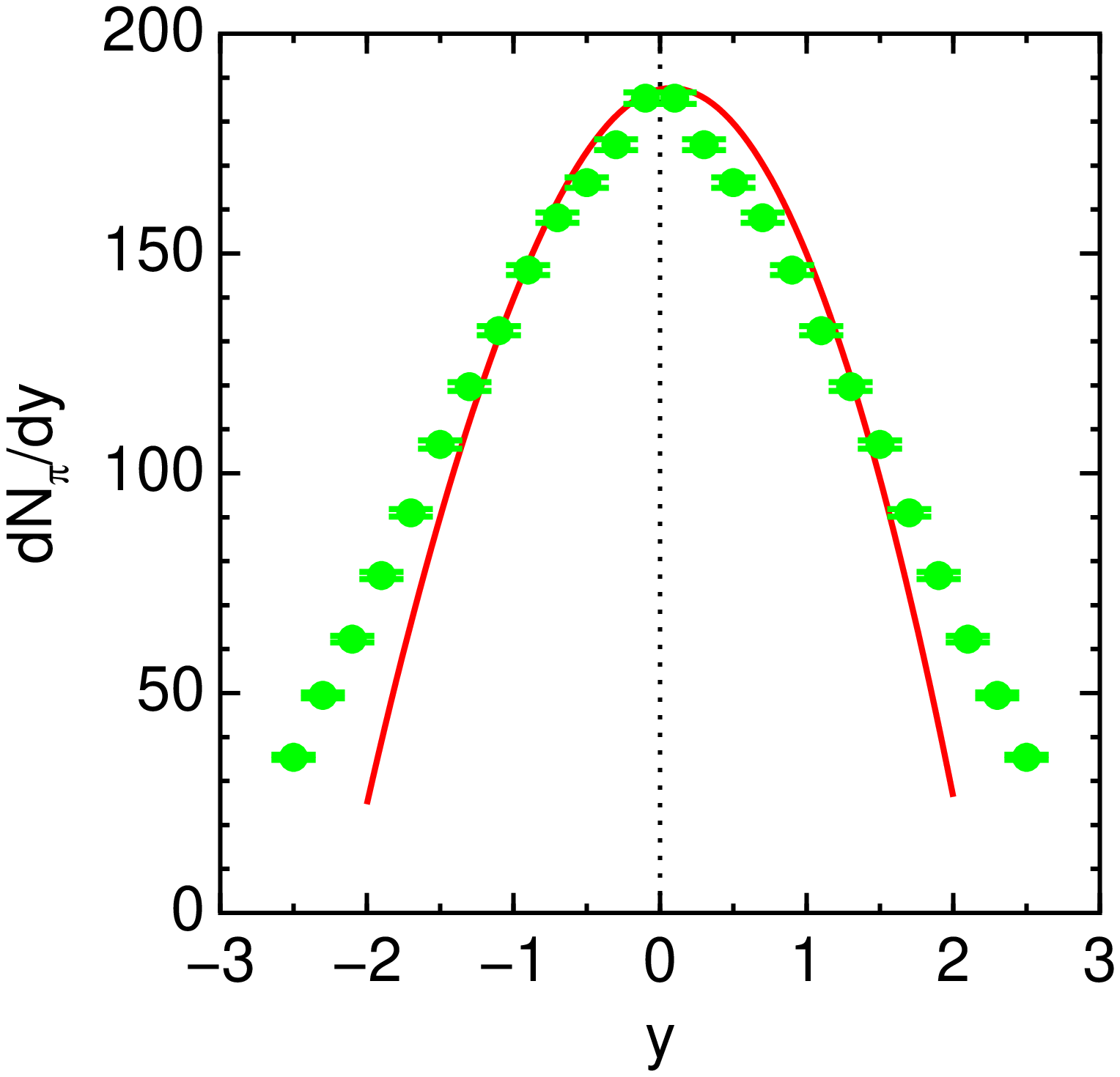} \hspace*{4mm}
    \includegraphics[width=6cm]{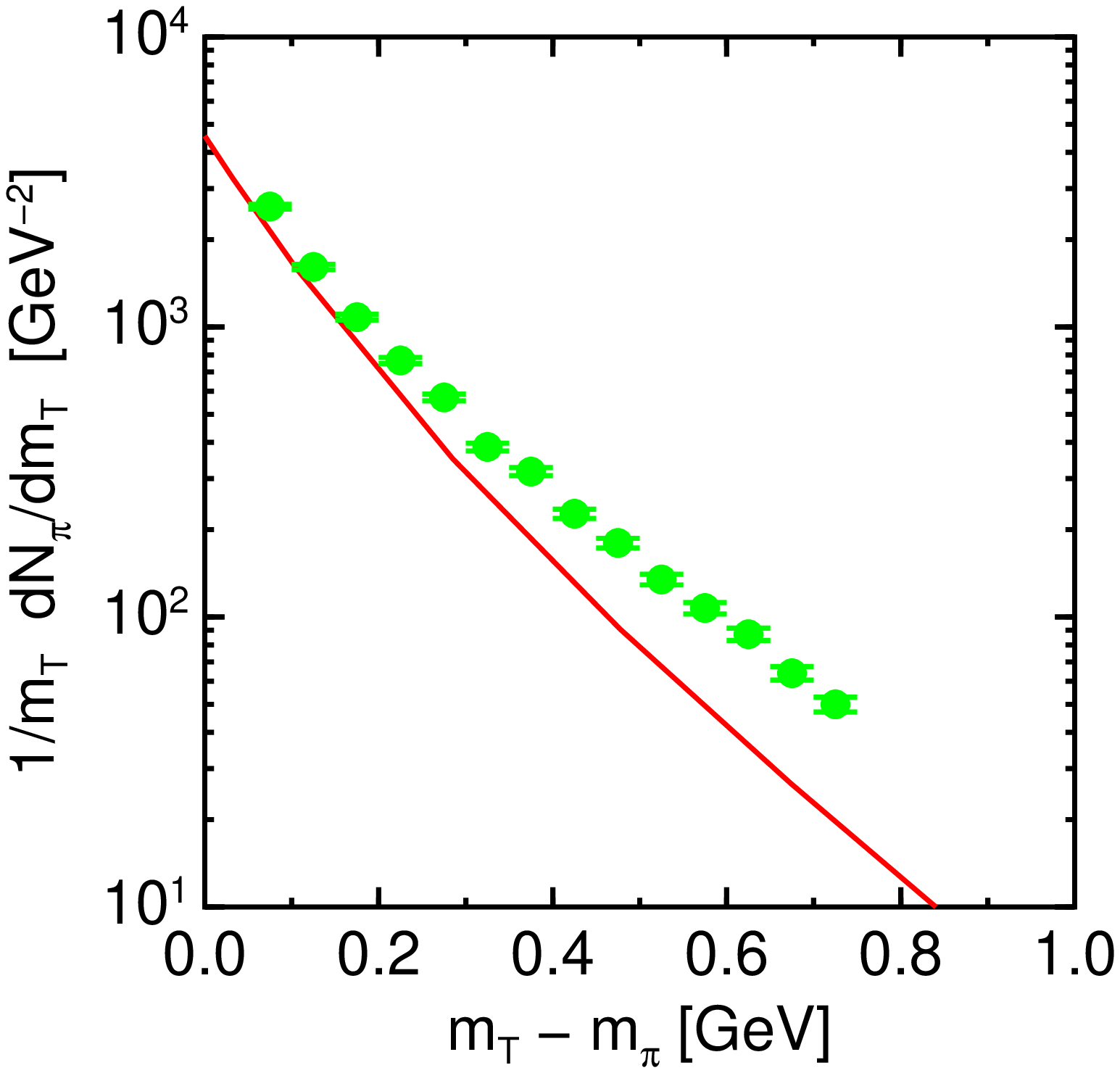}
 \caption{Rapidity (the left panel) and transverse mass (the right panel)
 distributions of $\pi^-$ meson from central Pb+Pb collision at
  $E_{lab}=158$ AGeV. Experimental points are taken from~\cite{AfNA49}.
  }
\label{fig:4}
\end{center}
\end{figure}

The full set of the freeze-out cells or the time-space freeze-out
surface, where hadrons are decoupled, defines all characteristics
of observable hadrons. To transform these frozen cells into
hadrons, we use the procedure described in~\cite{IRT05}.  The
freeze-out parameter $\varepsilon_{fr}$ is controlled by the pion
yield at high colliding energies (but weakly sensitive at moderate
energies, similarly to~\cite{IRT05}). Having in mind that the main
emphasis will be made on the $\pi\pi$ channel, in Fig.\ref{fig:4}
the rapidity and transverse mass spectra of observed pions are
shown for central Pb+Pb collisions at the top SPS energy. The
agreement between model calculations and data of the NA49
experiment is quite reasonable.

To find observable invariant mass spectra of dileptons, one should
integrate the emission rate ${d^8 N_{ee}(T(x),
\mu_B(x),M,\eta,q_T)} / {dt \ d^3 x  \ d^4q}$ over the whole
time-space evolution $\int dtd^3 x$ and  add the contribution from
the freeze-out surface (hadron cocktail). If we are interested in
the $e^+e^-$ invariant mass $M$ distributions in some
pseudo-rapidity window $\Delta \eta_{e\pm}$ (like in the CERES
experiment) and treat the  fireball evolution on average, then in
this case

\begin{eqnarray}
                \frac{d^2 N_{ee}}{d M d \eta} = \frac{M}{\Delta \eta_{e_{\pm}}}
\int d\eta \int V(t) dt \int_0^{2\pi} d\phi \int_0^\infty q_T \
dq_T \ \frac{d^5 N_{ee}(\bar{T}(t),\bar{\mu}_B(t),M,\eta,q_T)}{dt
\ d^4q} \ Acc,
                \label{rate_exp}
\end{eqnarray}
where the factor $Acc  \equiv Acc(M,\eta,q_T)$ takes into account
the experimental acceptance.

To simplify more our task, we  consider only the main emission
channel $\pi \pi \to l^+ l^-$. The dilepton emission rate
is~\cite{GK91}
\begin{equation}
\frac{d^5 N}{dt d^4q} = - {\cal L}(M) \ \frac{\alpha^2}{\pi^3 q^2}
\  f_B(q_0, \bar{T)} \ {\rm Im} \Pi_{em}(q,\bar{T},\bar{\mu}_b),
\label{rate1}
\end{equation}
where  the Bose distribution function is defined as
 $f_B(q_0, T) = (e^{q_0/\bar{T}} - 1)^{-1}$, the 4-momentum transfer
   $q^2=M^2=q_0^2-{\mathbf q}^2$
and the lepton kinematic factor is
\begin{equation}
{\cal L}(M) = \left(1+2\frac{m_l^2}{M^2}\right)
\sqrt{1-4\frac{m_l^2}{M^2}} \label{massive_factor1}
\end{equation}
with the lepton mass $m_l$.

We will consider  electro-magnetic current correlation function
$\Pi_{em}(q,\bar{T},\bar{\mu}_b)$ in two different scenarios:
(I) the finite temperature modification of pion gas properties~\cite{GK91};
(II) droping mass scenario \cite{RW00}.

\begin{figure}[t]
\begin{center}
  \includegraphics[width=6cm]{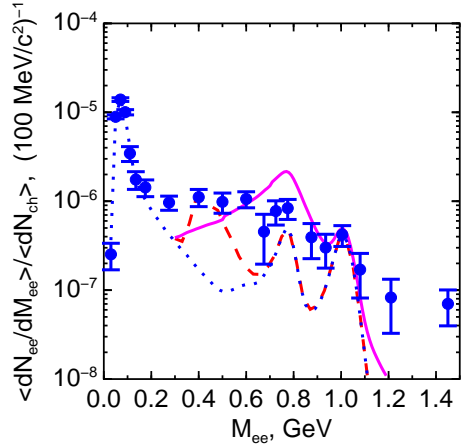}
  \caption{ Invariant mass distribution of dileptons from $8\%$ central Pb+Pb
  collisions at the beam energy 158 AGeV. Experimental points are
  from~\cite{CERES/NA45}. The solid and dashed
  curves are calculated using the finite temperature modification in pion gas
    and the $\rho$ dropping mass scenario
   (\ref{Tdep_mass}), respectively. The dotted line indicates an
   estimate of "hadron decay cocktail".}
\label{fig:6}
\end{center}
\end{figure}

Dilepton invariant mass spectra from $8\%$ centrality Pb+Pb
collisions at $E_{lab}=158$ AGeV are presented in Fig.\ref{fig:6}.
Experimental acceptance is taken into account for both hadron
cocktail and $\pi\pi$ annihilation channel. This channel is
calculated on the absolute basis without any fitting parameters.
As one can see, the 1st scenario slightly overestimates the
dilepton yield near the $\rho$ meson mass and underestimates it in
the $0.3\lsim M\lsim 0.6$ range though these values are
essentially above the hadron cocktail data. This overestimation is
explained mainly by simplified dynamics of dilepton production,
using average evolution  trajectories and rough approximations for
proceeding to hadronic phase. The depleted yield of low-mass
dileptons results from neglecting "sobar" contribution (nucleonic
loops) into the emission rate.

To simulate droping mass scenario, we modify the in-medium mass of
$\rho$-meson as commonly employed in the literature~:
\begin{equation}
m^*_{\rho} =  m_{\rho} (1-0.18 \cdot \bar{n}_B(t)/n_0)
\left(1-[\bar{T}(t)/T_c]^2\right)^{0.3}
 \label{Tdep_mass}
\end{equation}
and simultaneously apply the same modification to the vector
dominance coupling $m^2_\rho/g \to m^{*2}_\rho/g^*$. The first
factor in eq.(\ref{Tdep_mass}) was estimated by Hatsuda and
Lee~\cite{Hatsuda} according to the QCD sum rules. The temperature
dependence in eq.(\ref{Tdep_mass}) is motivated by the
T-dependence of the quark condensate.

The results for the dropping $\rho$ mass scenario are also plotted
in Fig.\ref{fig:6}. In this case, the low dilepton mass region
$0.2\lsim M\lsim 0.7$ is also populated but the free $\rho$ mass
region is essentially underestimated. The shape of the
$M$-distributions in these two scenarios is quite different. The
simplified dynamical assumptions mentioned above also influence
the $\rho$ mass scaling results. However, the large shift of the
$e^+e^-$ invariant mass spectrum towards low $M$ is mostly due to
using the T-dependent factor in eq.(\ref{Tdep_mass}), which seems
to be not very justified~\cite{BR04}.

Finally, inclusion of the full dynamics of heavy ion collisions in
evaluation of dilepton production is quite important. Comparison
of evolution within the proposed dynamical model with the popular
Bjorken regime shows essential differences which may manifest
themselves in dilepton observables. Full dynamical calculations
allow one to probe not only the shape of the $e^+e^-$ spectra from
$\pi\pi$ annihilation but also to estimate the absolute
contribution of this channel. Certainly, the simplifying dynamical
assumptions used should be overcome as well as other dilepton
production channels should be added, in particular those coming
from a quark phase. In this respect, it is of interest to note
that the mixed phase EoS, in principle, opens new possible
channels due to interaction between hadron and quark/gluon phases.
The hadron decay cocktail should also be calculated within the
same model. This work is now in progress.

\begin{ack}
We are grateful to Yu.B.~Ivanov and V.N.~Russkih for numerous
discussions  concerning hydrodynamic code and freeze-out
procedure. We thank S.~Yurevich for providing us with the hadronic
cocktail data for the CERES/NA45 experiment. One of us (VS) thanks
organizers of the QM2005 conference for making his participation
possible. This work was supported in part by DFG (project 436 RUS
113/558/0-3) and RFBR (grant 06-02-04001).
\end{ack}


\begin{thebibliography}{99}
\bibitem{CERES} G.~Agakishiev {\it et al.}, CERES Collaboration:  {\sl Phys. Rev.
Lett.} {\bf 75} (1995) 1272;  G.~Agakishiev {\it et al.}, CERES
Collaboration:  {\sl Phys. Lett.} {\bf B 422} (1998) 405
[arXiv:nucl-ex/9712008]; D.~Adamova {\it et al.}, CERES/NA45
Collaboration: {\sl Phys. Rev. Lett.} {\bf 91} (2003) 042301
[arXiv:nucl-ex/0209024].
%
\bibitem{CB99}W.~Cassing, E.~L.~Bratkovskaya: {\sl Phys. Rep.}
{\bf 308} (1999) 65.
%
\bibitem{RW00} R.~Rapp, J.~Wambach: {\sl Adv. Nucl. Phys.} {\bf 25}
(2000) 1.
%
\bibitem{BR91}G.~E.~Brown, M.~Rho: {\sl Phys. Rev Lett.} {\bf 66}
 (1991) 2720.
%
\bibitem{RW1}
  R.~Rapp, G.~Chanfray, J.~Wambach:
  %``Medium modifications of the rho meson at CERN SPS energies,''
  {\sl Phys. Rev. Lett.}  {\bf 76} (1996) 368
  [arXiv:hep-ph/9508353].
%
\bibitem{CERES/NA45} CERES/NA45 Collaboration: {\sl J. Phys.} {\bf G
30} (2004) S1007; {\sl J. Phys.} {\bf G 30} (2004) S2027;
S.~Yurevich (privite communication).
%
\bibitem{QGSM}
N.~S.~Amelin, K.~K.~Gudima, S.~Y.~Sivoklokov, V.~D.~Toneev:
  %``Further Development Of A Quark - Gluon String Model For Describing
  %High-Energy Collisions With A Nuclear Target,''
  {\sl Sov. J. Nucl. Phys.}  {\bf 52} (1990) 172
  [{\sl Yad. Fiz.}  {\bf 52} (1990) 272];
N.~S.~Amelin, E.~F.~Staubo, L.~P.~Csernai, V.~D.~Toneev,
K.~K.~Gudima:
%Strangeness Production in Proton and Heavy-Ion Collisions at 14.6 A GeV.
{\sl Phys. Rev.} {\bf C 44} (1991)  1541;
 V.~D.~Toneev, N.~S.~Amelin, K.~K.~Gudima, S.~Yu.~Sivoklokov:
 %Dynamics of  Relativistic  Heavy-Ion Collisions.
  {\sl Nucl. Phys.}  {\bf A 519} (1990) 463c.
%
\bibitem{ST05} V.~V.~Skokov, V.~D.~Toneev: {\it Hydrodynamics of an
expanding fireball}, JINR Preprint P2-2005-95 (2005) [accepted for
{\sl Yad. Fiz.}].
%
\bibitem{TFNFR04} V.~D.~Toneev, E.~G.~Nikonov,  B.~Friman, W.~N\"orenberg,
 K.~Redlich:
%Strangeness Production  in Nuclear Matter and Expansion Dynamics,
 {\sl Eur. Phys. J.} {\bf C 32} (2004) 399 [arXiv:hep-ph/0308088].
%
\bibitem{Dan} P.~Danielewicz, R.~Lacey, V.G.~Lynch: {\sl Science}
{\bf 298} (2002) 1592  [arXiv:nucl-th/0208016].
%
\bibitem{latEoS} Z.~Fodor:
%Lattice QCD results at finite temperature and density
{\sl Nucl. Phys.} {\bf A 715} (2003) 319;
  F.~Csikor, G.~I. ~Egri, Z.~Fodor, S.~D.~Katz, K.~K.~Szabo A.~I.~Toth,
%EQUATION OF STATE AT FINITE TEMPERATURE AND
%CHEMICAL POTENTIAL, LATTICE QCD RESULTS.
{\sl JHEP} {\bf 405} (2004) 46.
%
\bibitem{B83}J.~P.~Bjorken: {\sl Phys. Rev.} {\bf D 27} (1984) 140.
%
\bibitem{FNT98} B.~Friman, W.~N\"orenberg, V.~D.~Toneev:
%The Quark Condensate in Relativistic Nucleus-Nucleus Collisions,
{\sl Eur. Phys. J.} {\bf A 3} (1998) 165  [arXiv:nucl-th/9711065];
Bao-An~Li, C.~M.~Ko: {\sl Phys. Rev.} {\bf C 52} (1995)  2037.
%
\bibitem{IRT05} Yu.~B.~Ivanov, V.~N.~Russkikh, V.~D.~Toneev:
 {\it Relativistic Heavy-Ion Collisions within 3-Fluid Hydrodynamics:
Hadronic Scenario},  arXiv:nucl-th/0503088.
%
\bibitem{AfNA49}S.~V.~Afanasiev {\it et al.}, NA49 Collaboration:
{\sl Phys. Rev.} {\bf C 66} (2002) 054902
[arXiv:nucl-exp/0205002].
%
\bibitem{GK91} C.~Gale, J.I.~Kapusta: {\sl Nucl. Phys.} {\bf B 357} (1991) 65.
%
\bibitem{KKW96} F.~Klingl, N.~Kaiser, W.~Weise:
 %``Effective Lagrangian approach to vector mesons, their structure and
 %decays,''
 {\sl Z. Phys.} {\bf A 356} (1996) 193 [arXiv:hep-ph/9607431].
%
\bibitem{Hatsuda}
T.~Hatsuda, S.~H. ~Lee: {\sl Phys. Rev.} {\bf C 46} (1992) R34.
%
\bibitem{BR04} G.~E.~Brown, M.~Rho: {\sl Phys. Rept.} {\bf 396} (2004) 1.
%


\end{thebibliography}
\end{document}